\documentclass{article}
\usepackage{spconf,amsmath,graphicx}
\usepackage{multirow}
\usepackage{multicol}
\usepackage{epstopdf}
\usepackage{subfigure}
\title{Optimization of Occlusion-Inducing depth pixels in 3-D video Coding}
\name{Pan Gao$^{\star \dagger}$ \qquad Cagri Ozcinar$^{\dagger}$ \qquad Aljosa Smolic$^{\dagger}$ 
\thanks{This paper has been accepted at IEEE ICIP 2018, Athens, Greece. Personal use of this material is permitted. However, permission to reprint/republish this material for advertising or promotional purposes or for creating new collective works for resale or redistribution to servers or lists, or to reuse any copyrighted component of this work in other works must be obtained from the IEEE.}
}

\address{$^{\star}$ College of Computer Science and Technology, Nanjing University of Aeronautics and Astronautics \\
    $^{\dagger}$ V-SENSE Project, School of Computer Science and Statistics,  Trinity College Dublin
   }

\begin{document}
\ninept
\maketitle

\begin{abstract}
The optimization of occlusion-inducing depth pixels in depth map coding has received little attention in the literature, since their associated texture pixels are occluded in the synthesized view and their effect on the synthesized view is considered negligible. However, the occlusion-inducing depth pixels still need to consume the bits to be transmitted, and will induce geometry distortion that inherently exists in the synthesized view. In this paper, we propose an efficient depth map coding scheme specifically for the occlusion-inducing depth pixels by using allowable depth distortions. Firstly, we formulate a problem of minimizing the overall geometry distortion in the occlusion subject to the bit rate constraint, for which the depth distortion is properly adjusted within the set of allowable depth distortions that introduce the same disparity error as the initial depth distortion. Then, we propose a dynamic programming solution to find the optimal depth distortion vector for the occlusion. The proposed algorithm can improve the coding efficiency without alteration of the occlusion order. Simulation results confirm the performance improvement compared to other existing algorithms.
\end{abstract}
\begin{keywords}
Occlusion, depth map coding, allowable depth error, optimization
\end{keywords}
%

\section{Introduction}
\label{sec:intro}
Depth map has been a very important signal used in emerging 3-D video due to the capability of synthesizing the virtual views that are not captured or transmitted at the sender \cite{3D-Video-Compression-Overview}.  In view synthesis, the main role of the depth map is to provide the geometrical information that determines how far the associated texture pixel in real view is shifted to  the virtual view. In practice, the foreground objects may be shifted further than the background, and thus it is possible that some occlusions occur after view synthesis \cite{Fehn}. Since the associated texture pixels are already occluded there, the effect of occlusion-inducing depth pixels on the synthesized view is usually neglected or  considered insignificant in depth map coding and optimization.

An example of this consideration can be found in \cite{Kim_update}, where, by considering that every reference texture pixels can be mapped to a specific pixel in the synthesized view, an autoregressive model was developed to characterize the rendered view distortion at the block level, which is then used in mode selection for efficient depth map coding. Similarly, Yuan \emph{et al.}\cite{Yuan_CSVT} proposed a joint bit allocation scheme for texture and depth map coding, which is based on a  derived  polynomial synthesis distortion model. Taking into account texture image characteristics, Fang \emph{et al.} \cite{Lu_TIP} proposed to estimate the rendering distortion using an approach that combines frequency domain analysis with power spectral density and spatial domain analysis with local gradient information. In all the algorithms discussed, the reference texture are assumed to be bijectively mapped to the synthesized view without occlusion.
With the observation that a reference texture pixel may not have information to contribute to  a specific synthesized pixel, Velisavljevic \emph{et al.} \cite{Velisavljevic} developed a cubic synthesized view distortion model as a function of the view's location for multiview image compression, and experimentally demonstrated that the number of occluded pixels in the synthesized view is linearly proportional to the distance between the reference view and synthesized view. By further analyzing different possible overlapping/disocclusion scenarios and the effects due to different extents of depth error, Fang \emph{et al.} \cite{Fang_TIP2} extended the cubic model to an analytic quintic model, which mathematically combines the quadratic/biquadratic models for view synthesis distortion and the linear models for the probability  under different defined scenarios. Although these two algorithms hand the occlusion in the prediction of view synthesis distortion, they explicitly skip the distortion of the synthesized pixels that are occluded, and only calculate the distortion of the winning pixel as the overall distortion of the occlusion. As a result, the  estimated distortion may not be appropriate  for occlusion-specific depth coding optimization,  since there are still some distortions occurred by the process of shifting the occluded pixels from original positions to the destination positions, although the associated texture distortions  are zero intuitively.

In this paper, we focus on addressing the problem of how to optimize the depth pixels that induce the occlusion in depth map coding in a rate-distortion sense. In general, the optimization of occlusion-inducing depth pixels is a lot challenging since the occlusion order can be easily modified by the optimization algorithm. In this work, we resort to the \emph{allowable depth distortion} redundancy for depth coding in occlusion scenario. The allowable depth distortion here refers to the depth distortion that generates the same amount of view synthesis distortion as the initial depth distortion induced by quantization.
It is firstly proposed in  \cite{YinZhao}, where a depth no-synthesis-error (D-NOSE) model for view synthesis is derived based on the observation that multiple depth levels may correspond to the same disparity due to the disparity rounding function. Later, the allowable depth distortion model is enhanced in \cite{YunZhang} by considering that there also exists an allowable depth level change range for the depth error that leads to one non-zero disparity error.
Since allowable depth distortions can keep the view synthesis distortion unchanged, the occlusion order in which the quantized depth levels proceed can  be preserved if the initial depth distortions of all the pixels are changed to their own allowable depth distortions. In addition, different allowable depth distortions may lead to different overall view synthesis costs, which means that the selection of allowable depth distortion has the potential to improve the coding efficiency.

However, optimization of occlusion-inducing depth pixels using allowable depth distortions still faces major challenges since there will be distortion  dependency introduced by warping competition between depth pixels involved in the occlusion. To overcome the interdependence between depth pixels in occlusion, we firstly model the occlusion distortion as the summation of the view synthesis distortions of a group of pixels that are involved in the occlusion. Then, we formulate the optimization problem as optimally selecting the possible allowable depth error from the derived range for each pixel such that the overall distortion involved in the occlusion is minimized. Finally,  we propose a dynamic programming solution to efficiently find the vector of allowable depth level changes.

The rest of this paper is organized as follows. In Section 2, we provide a review of how to derive allowable depth errors for a depth level with a quantized error. Section 3 develops a Lagrangian-based method specifically for optimization of occlusion-inducing depth pixels by using allowable depth errors. Experimental results are presented and discussed in Section 4. Finally, concluding remarks are presented in Section 5.

\section{Preliminaries}
In this section, we review how to determine the allowable depth distortion range, which inherently exists for each possible depth error of each depth pixel. In view synthesis,  the depth distortion of a pixel may generate the disparity error. However, due to the significantly less number of disparity levels compared to the number of available depth levels, several depth errors for a certain pixel may correspond to the same disparity error. The depth distortion that induces the same disparity error as the current depth distortion is thus called allowable depth distortion, which can be derived depending on whether the corresponding disparity error is zero. 

Assume that $v_i$ is the depth level of a pixel $i$  and $D(\cdot)$ is the disparity function of the depth level. Then, the disparity that is induced by $v_i$ can be expressed as follows \cite{depth-measurement}
\begin{equation}
D(v_i)  = fl(C_1  \cdot v_i  + C_2 )
\end{equation}
where $f$ is  the focal length, $l$ is the baseline distance between the reference and virtual views, ${C_1} = 1/255(1/{Z_{{\rm{near}}}} - 1/{Z_{{\rm{far}}}})$, and
${C_2} = 1/{Z_{{\rm{far}}}}$. ${Z_{{\rm{near}}} }$ and ${Z_{{\rm{far}}} }$ denote the nearest and farthest depth values of the scene, respectively. Generally, the disparity needs to be rounded to a $1/N$ sub-pixel sampling position with the following rounding function \cite{YinZhao}
\begin{equation}\label{disparity_rounding}
R(D(v_i )) = {\left\lceil {(D(v_i ) - o)N} \right\rceil }/N
\end{equation}
where $o$ represents the offset error and determines the decision level of the rounding process. The value of $o$ is
$0 < o \le 1/N$.

Assume now pixel $i$ has a depth level change of $\Delta v_i$ caused by compression, and thus the resulting depth level $v_i+\Delta v_i$ corresponds to another disparity for pixel $i$. Therefore, the disparity error  due to $\Delta v_i$  can be written as
$R(D(v_i  + \Delta v_i )) - R(D(v_i ))$. Due to limited number of disparity levels, there exists a range of depth level change that makes the associated disparity error equal to zero. Based on \eqref{disparity_rounding}, the sufficient condition for depth level $v_i$ of no disparity difference introduced in pixel mapping is that  $v_i+\Delta v_i$ corresponds to the same rounded disparity as what $v_i$ generates. In light of this, we have
\begin{equation}\label{sufficient_condition}
 - \left( {\frac{1}{N} - o} \right) \le D(v_i  + \Delta v_i ) - R(D(v_i )) < o
\end{equation}

After some substitutions and rearrangements for \eqref{sufficient_condition}, we can get the allowable depth level variation range $\Delta V_i$ for depth level $v_i$ for the case when the associated disparity error is zero. For simplicity, we use $\Delta v_i^ -$ and $\Delta v_i^ +$ to represent the lower bound and upper bound of $\Delta v_i$, respectively. In this case, the zero-disparity-error  interval of allowable depth level change can be defined as
$\Delta v_i  \in \Delta V_i  = [\Delta v_i^ -  ,\Delta v_i^ +  ]$.

When there is a depth error $\Delta v_i^k$ that induces a non-zero disparity error, there also exists an allowable depth error range within which the depth error generates the same non-zero disparity error as $\Delta v_i^k$. Based on the derivation in \cite{YunZhang}, the interval of allowable depth level change around $\Delta v_i^k$ can be modeled as the zero-disparity-error allowable depth error range of the initial depth level scaled by adding the associated depth error to its lower and upper bounds, i.e., $[ \Delta v_i^k+\Delta v_i^ -  ,\Delta v_i^k+ \Delta v_i^ +  ]$, where $[\Delta v_i^ -,\Delta v_i^ +  ]$ is the range of allowable depth level change of pixel $i$ relative to initial depth level $v_i$ in the zero disparity error case.

\vspace{-2mm}
\section{Coding Optimization of Occlusion-Inducing Depth Pixels}
\vspace{-2mm}
In this section, we elaborate on how to jointly optimize the occlusion-inducing depth pixels using allowable depth errors in a rate-distortion optimal manner.
When an occlusion occurs after view synthesis using the reconstructed texture and depth, there are multiple texture pixels in the reference view that are mapped to a same position in the virtual view, and the texture pixel with the largest quantized depth level is chosen as the final synthesized pixel. Let $N$ be the total number of pixels involved in this occlusion. Assume now the pixel $j$ is the final wining pixel, which means that the depth levels of all the previous pixels are smaller than the depth level of the wining pixel, i.e.,
$v_i  + \Delta v_i  < v_j  + \Delta v_j :i \in [1,j - 1]$, and the depth levels of all the subsequent pixels are smaller than or equal to that of the wining pixel, i.e., $v_i  + \Delta v_i  \leq v_j  + \Delta v_j :i \in [j+1,N]$.
Recall that $\Delta v_j $ is the quantized depth error, which can be optimized based on allowable depth errors for coding efficiency improvement.
Let $P(\Delta v_j )$ be the probability of the depth pixel $j$ choosing the possible depth change
$\Delta v_j$ from allowable depth error range, and ${\bar d_j (\Delta v_j )}$ be the associated view synthesis distortion caused by the particular depth error $\Delta v_j$. Therefore, the final view synthesis distortion of choosing $\Delta v_j$ for pixel $j$ can be represented as
\begin{equation}\label{Formulation_Single_Pixel}
\begin{array}{lll}
 D(\Delta v_j ) &=& P(\Delta v_j ) \times \prod\limits_{i = 1}^{j - 1} {\mathop P\limits_{v_i  + \Delta v_i  < v_j  + \Delta v_j } (\Delta v_i )}  \\
 && \times \prod\limits_{i = j + 1}^N {\mathop P\limits_{v_i  + \Delta v_i  \le v_j  + \Delta v_j } (\Delta v_i )}  \times  {\bar d_j (\Delta v_j )}  \\
 \end{array}
\end{equation}

In \eqref{Formulation_Single_Pixel}, ${\bar d_j (\Delta v_j )}$ can be estimated from the rendering position error that is induced by depth error $\Delta v_i $ using the methods in \cite{depth-measurement}, \cite{Partial-DIBR}. As can be observed from \eqref{Formulation_Single_Pixel}, the selection of the depth changes of other pixels could affect the choice of that of the current pixel, and thus the depth error selection of the current pixel should be jointly considered with other pixels. In the following, considering the correlation between the pixels involved in the warping competition process, we formulate a view synthesis distortion criterion for all the pixels. For ease of derivation, we rearrange the pixels in a depth-level-monotonic-increasing order, with the wining pixel being the last one. Let
${\bf{\Delta v}} = [\Delta v_1, \Delta v_2,  \cdots, \Delta v_N ]$ be a vector representing the depth level changes assigned to the $N$ pixels. The total view synthesis cost caused by the selection of the depth change within the allowable depth range for a given total bit rate $R_c$ can be formulated as follows
\begin{equation}\label{rate-distortion-cost}
\begin{array}{lll}
 J({\bf{\Delta v}}) &=& \sum\limits_{z = 1}^N \Bigg\{ {P(\Delta v_1 )\left( {\prod\limits_{j = 2}^{z - 1} {\mathop P\limits_{v_{j - 1}  + \Delta v_{j - 1}  \le v_j  + \Delta v_j } (\Delta v_j )} } \right)}  \\
 && \times \mathop P\limits_{v_{z - 1}  + \Delta v_{z - 1}  < v_z  + \Delta v_z } (\Delta v_z ) \times \bar d_z (\Delta v_z ) \Bigg\}\\
 &&+ \lambda \sum\limits_{z = 1}^N {R_z (\Delta v_z )}  \\
 \end{array}
\end{equation}
where $R_z (\Delta v_z )$ denotes the bit rate for pixel $z$ when the pixel value $v_z$ is changed to $v_z+\Delta v_z$. $\lambda$ is called the Lagrangian multiplier. Then, the depth coding optimization using allowable depth distortions for the occlusion scenario is to search $({\bf{\Delta v}})^*$ which minimizes $J({\bf{\Delta v}})$, i.e.,
\begin{equation}\label{argmin}
({\bf{\Delta v}})^*  = \arg \mathop {\min }\limits_{{\bf{\Delta v}}} J({\bf{\Delta v}})
\end{equation}

If $({\bf{\Delta v}})^*$ leads to that $\sum\limits_{z = 1}^N {R_z (\Delta v_z )}$ happens to $R_c$, then $({\bf{\Delta v}})^*$ is also an optimal solution to the constrained problem of minimizing the total view synthesis distortion subject to a given maximum bit rate of  $R_c$. It is well-known that when $\lambda$ sweeps from zero to infinity, the solution to \eqref{argmin} traces out the convex hull of the rate distortion curve, which is a nonincreasing function. Hence, the bisection method in \cite{bisection_Anea} can be used to find the optimal $\lambda$.

Therefore, the task at hand is  to find the optimal solution to the problem expressed in \eqref{argmin}.  One possible way to accomplish this task is by exhaustive search. However,  enumerating all possible ${\bf{\Delta v}}$ and substituting into \eqref{argmin} will render the solution to the optimization of depth pixels intractable. In this paper, we propose a dynamic programming algorithm to find the optimal combination of depth errors for the occlusion-inducing depth pixels.

As demonstrated in \cite{YinZhao} and \cite{YunZhang}, if the depth distortions of all the pixels in the occlusion are varied within their own  allowable depth error range, the occlusion order formed by the quantized depth levels can be preserved.
Therefore, the subscript underneath $P(\Delta v_j )$ indicating the constraint of monotonic increase of depth levels can be dropped for convenience, which leads to \eqref{rate-distortion-cost} as
\begin{equation}\label{simplified_RD_cost}
J({\bf{\Delta v}}) = \sum\limits_{z = 1}^N {\left\{ {\left( {\prod\limits_{j = 1}^z {P(\Delta v_z )} } \right) \times \bar d_z (\Delta v_z ) + \lambda R_z (\Delta v_z )} \right\}}
\end{equation}

Based on \eqref{simplified_RD_cost}, for the purpose of forward dynamic programming, the cost function at state $\Delta v_k$ of stage $k$ is defined as
\begin{equation}
g_k (\Delta v_k ) = P(\Delta v_k )\bar d_k (\Delta v_k ) + \lambda R_k (\Delta v_k )
\end{equation}
where stage $k$ corresponds to the $k$th pixel in the occlusion ($1 \le k \le N$), and state $\Delta v_k$ is selected from the set of available depth level changes of stage $k$, i.e., $\Delta V_k$.

Then, the cost-to-go function at  state $\Delta v_k$ of stage $k$ can be written as
\begin{equation}
J_1 (\Delta v_1 ) = g_1 (\Delta v_1 )\quad ({\rm{when }}~~k = 1)
\end{equation}
and when, $2 \le k \le N$, as shown in \eqref{cost-to-go-function} below,
\begin{equation}\label{cost-to-go-function}
\begin{array}{lll}
J_k (\Delta v_k )
 &=&\min \limits_{\Delta v_{k - 1}  \in \Delta V_{k - 1} } \Bigg\{ J_{k - 1} (\Delta v_{k - 1} )\\
 &&+  \left( {\prod\limits_{j = 1}^{k - 2} {P(\Delta v_j^* )} } \right)
 P(\Delta v_{k - 1} )[g_k (\Delta v_k ) - \lambda R_k (\Delta v_k )]\\
 &&+ \lambda R_k (\Delta v_k )\Bigg\}
\end{array}
\end{equation}
where
\begin{equation}\label{optimal_path}
\Delta v_j^*  = \left\{ \begin{array}{l}
 \mathop {\arg \min }\limits_{\Delta v_{k - 2}  \in \Delta V_{k - 2} } J_{k - 1} (\Delta v_{k - 1} ),\quad j = k - 2 \\
 \mathop {\arg \min }\limits_{\Delta v_j  \in \Delta V_j } J_{j + 1} (\Delta v_{j + 1}^* ),\quad 1 \le j \le k - 3 \\
 \end{array} \right.
\end{equation}

For a given $\lambda$, the proposed dynamic programming solution proceeds from the first pixel to the wining pixel. For each state of stage $k$, it evaluates all the paths that lead to the state from any admissible state in the previous stage $k-1$, and store the one that produces the minimum Lagrangian cost. At the last stage, the minimal cost associated with the state becomes the optimal cost, and the path that leads to this state from the first stage given by \eqref{optimal_path} determines the optimal depth level change vector ${\bf{\Delta v}}$.

\section{Experimental Results and Discussion}
In the experiment, we implement the proposed occlusion-aware depth map optimization algorithm on the 3D-HEVC reference software HTM-16.0 \cite{3D-HTM}, and the VSRS-1D-Fast view synthesis software included in HTM 16.0 is used to render the  intermediate virtual views. The standard multi-view video sequences ``BookArrival", ``Lovebird1" and ``Undo\_Dancer" are chosen for our simulations. Note that the first two sequences have a resolution of 1024$\times$768, while the resolution of the rest sequence is 1920$\times$1088.
For each  sequence, each view is encoded with a group of pictures (GOP) size of 8 frames,  and the intra period is 24.  For 3-D video coding, the encoder uses variable block-size motion and disparity estimation, with a search range of 64 pels. Three QP combinations for texture and depth are considered: (25;34), (30;39), and (35;42).
The virtual views are  generated with half-pel precision and symmetric rounding.  It should be noted that the experimental setup is in accordance with the Common Test Condition of the Joint Collaborative Team for 3DV \cite{CTC}. The depth error is assumed to follow the Gaussian distribution.

For performance evaluation, Bj{\o}ntegaard Delta Bit Rate (BDBR) \cite{Bjontegaard} is used for objective video quality assessment, which is measured by the total bit rate of the texture and depth along with the average peak signal-to-noise ratio (PSNR) of the synthesized views.  In the comparison, the original 3D-HTM without allowable depth distortion consideration is used as the anchor. The  D-NOSE-based depth  coding algorithm proposed in \cite{YinZhao} and the  allowable depth distortion controlled depth coding algorithm developed in \cite{YunZhang} are employed as competing approaches, which are referred to as D-NOSE\_DC and ADD\_DC, respectively. The proposed allowable depth distortion based occlusion-aware depth map coding scheme in this work is denoted by ADD\_ODC for brevity. It should be noted that, in ADD\_ODC, the depth pixels that are not involved in occlusion are optimized in an independent manner, i.e., each pixel selects the allowable depth distortion within its own range to minimize the view synthesis cost.

To examine the effect of the occlusion on the optimization performance, we test the above algorithms by using different proportions of occluded pixels in the synthesized view. As has been demonstrated in \cite{Velisavljevic} and \cite{Fang_TIP2}, the number of occluded pixels in the synthesized view is linearly proportional to the distance between the reference view and synthesized view. Inspired by this, we manually vary the distance of the virtual view with respect to the reference view to generate different proportions of occluded pixels. Specifically, in this test, the views 7, 9 and 4 are assumed to be the virtual views for ``Lovebird1", ``BookArrival", and ``Undo\_Dancer", respectively, which need to be synthesized. For the coded views, we employ three different pairs of real camera-captured views for encoding for each sequence.
The specific selected captured views and the related results are shown in Table \ref{TABLE_BITRATE_OCCLUSION_CHANGE}. As can be observed, when the relative distance increases, the performances of D-NOSE\_DC and ADD\_DC are significantly degraded, and the performance gaps between the reference algorithms and ADD\_ODC become larger. This is mostly due to the fact that, both D-NOSE\_DC and ADD\_DC do not optimize the pixels that are involved in occlusion, which in fact can contribute to the bit rate required to be transmitted and overall view synthesis distortion. In contrast, ADD\_ODC minimizes the overall view synthesis distortion of occlusion by optimally allocating the allowable depth distortions (or bits) between the associated pixels.
{
\begin{table}[!htbp]
	\vspace{-2mm}
\renewcommand{\arraystretch}{1.3}
\begin{footnotesize}
\caption{\protect\\BDBR  comparison of the schemes D-NOSE\_DC, ADD\_DC and ADD\_ODC for a variety of test sequences  with three different pairs of input views coded for each sequence. The views 7, 9 and 4 are the virtual intermediate views for Lovebird1, BookArrival, and Undo\_Dancer, respectively.} \label{TABLE_BITRATE_OCCLUSION_CHANGE}
\end{footnotesize}
\centering
\begin{footnotesize}
\begin{tabular}{c|c|c|c|c}
	\hline
\multicolumn{1}{c|}{\multirow {2}{*}{Test sequence}}&\multicolumn{1}{c|}{\multirow {2}{*}{Input views}}& \multicolumn{3}{c}{BDBR (\%)}\\\cline{3-5}

\multicolumn{1}{c|}{}&
\multicolumn{1}{c|}{}
&\multicolumn{1}{c|}{\scriptsize{D-NOSE\_DC}}
&\multicolumn{1}{c|}{\scriptsize{ADD\_DC}}
&\multicolumn{1}{c}{\scriptsize{ADD\_ODC}}
\\\cline{1-5}

\multicolumn{1}{c|}{\multirow {3}{*}{Lovebird1}}&\multicolumn{1}{c|}{6-8}&\multicolumn{1}{c|}{$-$5.65}
&\multicolumn{1}{c|}{$-$10.35}&\multicolumn{1}{c}{$-$14.32}\\\cline{2-5}

\multicolumn{1}{c|}{}&\multicolumn{1}{c|}{5-9}&\multicolumn{1}{c|}{$-$3.26}
&\multicolumn{1}{c|}{$-$8.46}&\multicolumn{1}{c}{$-$13.48}\\\cline{2-5}

\multicolumn{1}{c|}{}&\multicolumn{1}{c|}{4-10}&\multicolumn{1}{c|}{$-$2.59}
&\multicolumn{1}{c|}{$-$6.11}&\multicolumn{1}{c}{$-$13.24}\\\cline{1-5}

\multicolumn{1}{c|}{\multirow {3}{*}{BookArrival}}&\multicolumn{1}{c|}{8-10}&\multicolumn{1}{c|}{$-$6.39}
&\multicolumn{1}{c|}{$-$12.65}&\multicolumn{1}{c}{$-$16.26}\\\cline{2-5}

\multicolumn{1}{c|}{}&\multicolumn{1}{c|}{7-11}&\multicolumn{1}{c|}{$-$4.23}
&\multicolumn{1}{c|}{$-$10.15}&\multicolumn{1}{c}{$-$15.10}\\\cline{2-5}

\multicolumn{1}{c|}{}&\multicolumn{1}{c|}{6-12}&\multicolumn{1}{c|}{$-$3.72}
&\multicolumn{1}{c|}{$-$7.19}&\multicolumn{1}{c}{$-$16.28}\\\cline{1-5}

\multicolumn{1}{c|}{\multirow {3}{*}{Undo\_Dancer}}&\multicolumn{1}{c|}{3-6}&\multicolumn{1}{c|}{$-$7.69}
&\multicolumn{1}{c|}{$-$11.21}&\multicolumn{1}{c}{$-$17.06}\\\cline{2-5}

\multicolumn{1}{c|}{}&\multicolumn{1}{c|}{2-7}&\multicolumn{1}{c|}{$-$5.21}
&\multicolumn{1}{c|}{$-$8.34}&\multicolumn{1}{c}{$-$17.15}\\\cline{2-5}

\multicolumn{1}{c|}{}&\multicolumn{1}{c|}{1-8}&\multicolumn{1}{c|}{$-$4.12}
&\multicolumn{1}{c|}{$-$6.63}&\multicolumn{1}{c}{$-$16.45}\\\cline{1-5}

\hline

\end{tabular}
\end{footnotesize}
\end{table}
}

\begin{figure}[!htbp]
	\centering
	\subfigure[]{
		\label{Fig.sub.1}
		\includegraphics[width=0.23\textwidth]{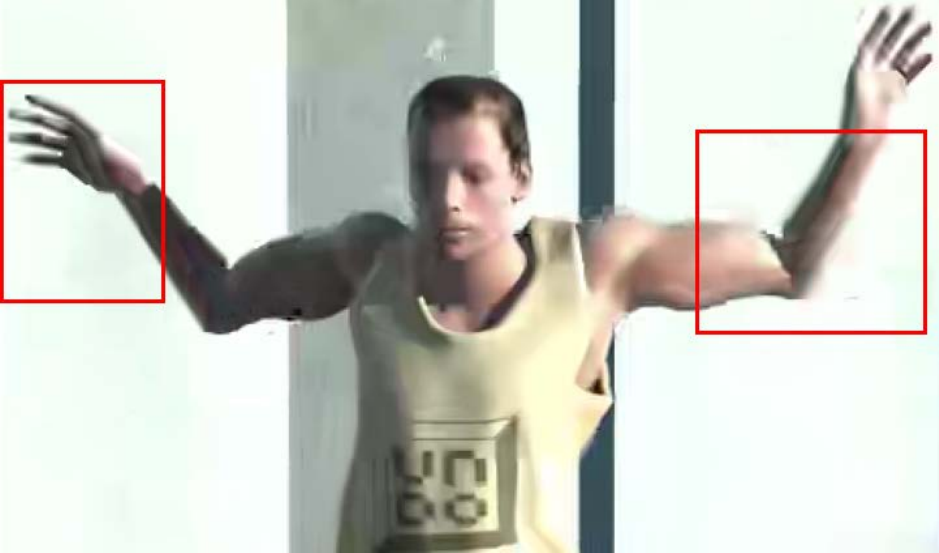}}
	\subfigure[]{
		\label{Fig.sub.1}
		\includegraphics[width=0.23\textwidth]{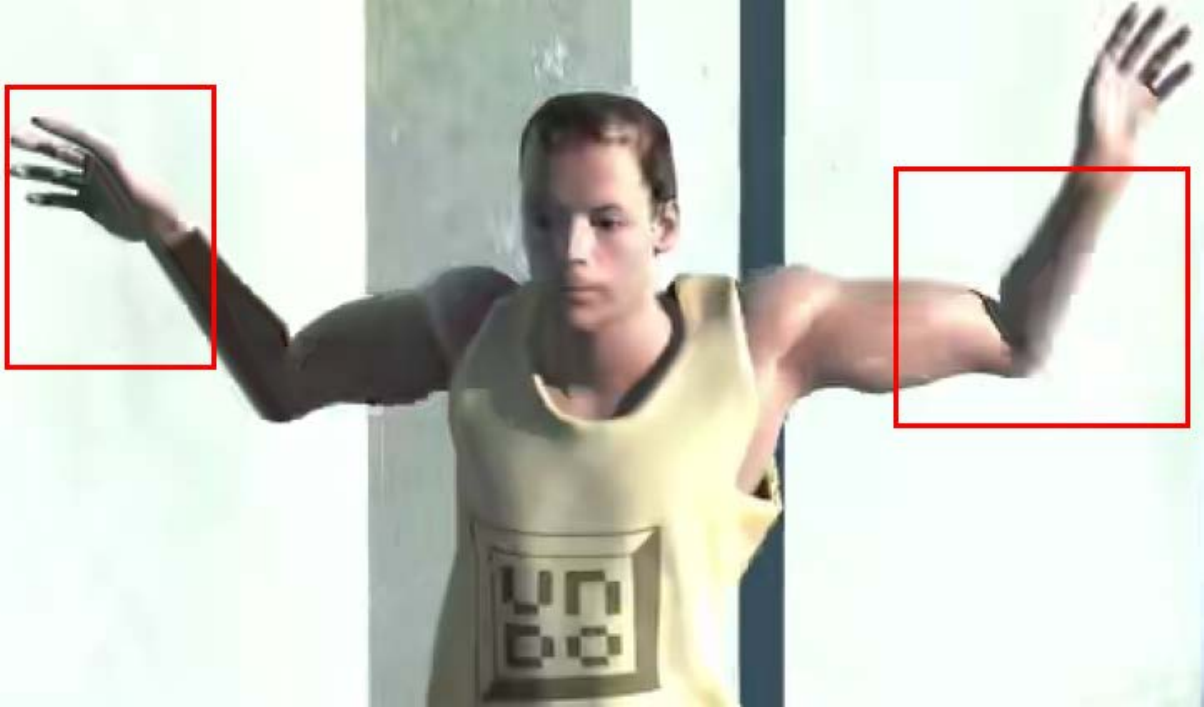}}
		\subfigure[]{
		\label{Fig.sub.1}
		\includegraphics[width=0.23\textwidth]{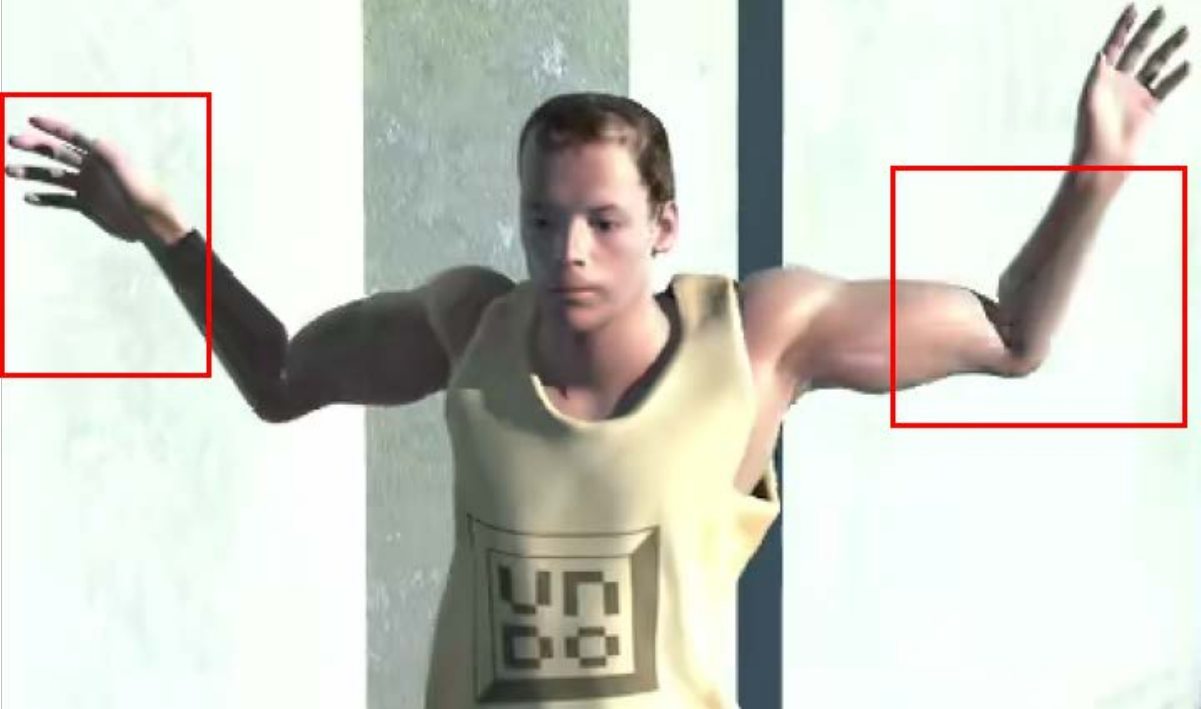}}
	\subfigure[]{
		\label{Fig.sub.1}
		\includegraphics[width=0.23\textwidth]{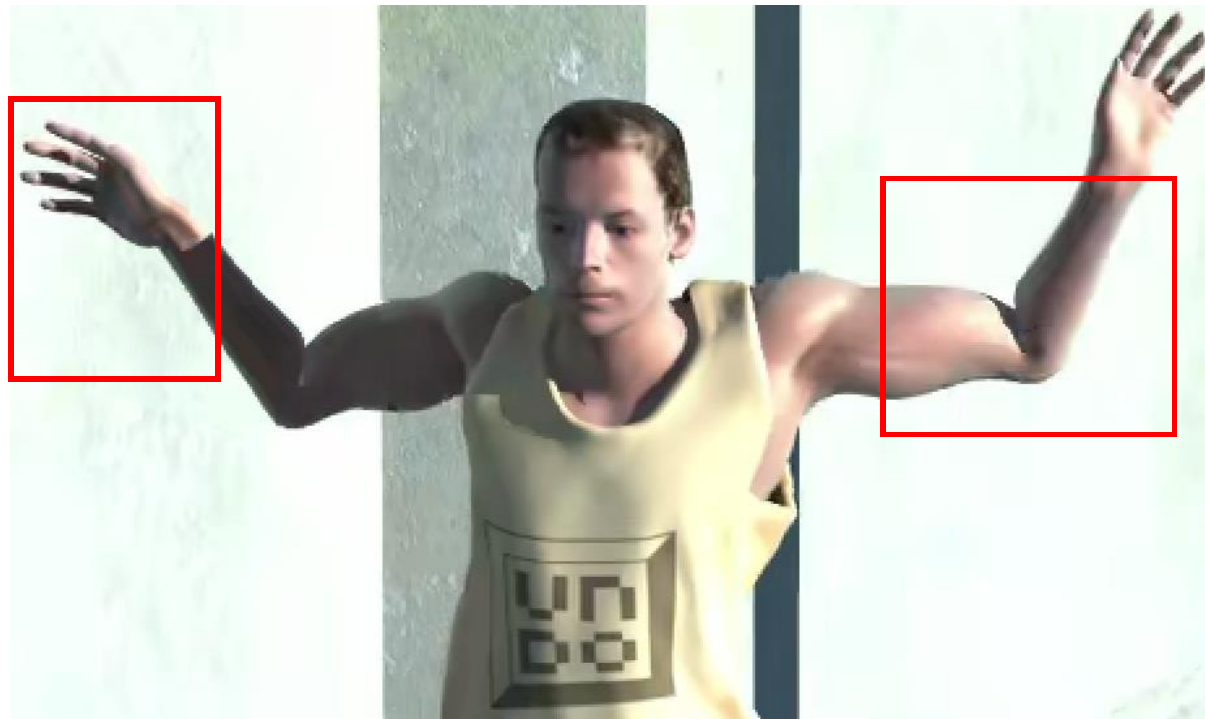}}
	\caption{Subjective comparison of the synthesized views of the Undo\_Dancer sequence between the proposed ADD\_ODC algorithm and the reference algorithms. (a) Synthesized view rendered with anchor-coded depth images. (b) Synthesized view rendered with D-NOSE\_DC-optimized depth images. (c) Synthesized view rendered with ADD\_DC-optimized depth images. (d) Synthesized view rendered with ADD\_ODC-optimized depth images.}
	\label{visual_comparison}
\end{figure}

For subjective quality evaluation, Fig. \ref{visual_comparison} shows the synthesized views for the Undo\_Dancer sequence using encoding schemes of D-NOSE\_DC, ADD\_DC, and ADD\_ODC. For better comparison, we also include the synthesized view rendered by the anchor-scheme-coded texture and depth as benchmark. As can be observed, all the test algorithms can improve the rendered view quality visually compared to the anchor, and ADD\_ODC achieves the largest subjective quality gain. Especially, we can also observe that ADD\_ODC generally yields better synthesis quality around the area of object edges where occlusion frequently occurs.

\section{Conclusions}

In this paper, we proposed an algorithm that jointly optimizes the occlusion-inducing depth pixels by using allowable depth errors in 3-D video coding. Firstly, considering the interdependence between the pixels caused by warping competition, we model the synthesis distortion for occlusion as the summation of the view synthesis distortions of all the pixels involved. Then, joint optimization of depth pixels is derived using a Lagrange multiplier method, where the depth distortion for each pixel is dynamically adjusted within its allowable depth error interval. Finally, a dynamic programming solution is proposed to find the optimal depth level change vector. Experimental results demonstrate the effectiveness of the proposed allowable-depth-error-controlled algorithm in optimizing the occlusion-related depth pixels.

\vspace{-2mm}


\end{document}